# A Multi-stack Power-to-Hydrogen Load Control Framework for the Power Factor-Constrained Integration in Volatile Peak Shaving Conditions

Jiarong Li

*Abstract*—Large-scale power-to-hydrogen (P2H) systems formed by multi-stack are potentially powerful peak-shaving resources of power systems. However, due to the research gap in connecting the grid-side performance with the inherent operation control, the continuous operation of P2H loads is limited by the PF assessment under volatile conditions when integrating into the grid. This paper first fills the gap in proposing the analytical models of active and reactive power of P2H loads with a typical power converter interface topology. On this basis, the all-condition PF characteristics of multi-stack P2H loads are captured as functions of unified current and temperature control variables. Then, a PF-constrained multi-timescale control framework is constructed to evaluate flexibility, PF, production, and security comprehensively. A two-level nexus, including a model-based hour-ahead robust model predictive controller and a rule-based real-time increment correction algorithm, is proposed to guarantee the control accuracy and tractability. Case studies verify an intrinsic control tradeoff between PF and production, resulting in an unequal-split allocation strategy compared to the traditional production-oriented control. The significance of the extended PF and security dimensions is verified to improve the flexibility. Furthermore, five typical operating modes respectively corresponding to low, medium, and high load levels at the cluster level are concluded for industrial application.

## NOMENCLATURE

The main variables and input parameters in this paper are listed below; other symbols are defined as required.

### A. Abbreviations

| | |
|---|---|
| IEA | International Energy Agency |
| AEL | Alkaline electrolyzer |
| AC | Alternate current |
| DC | Direct current |
| HX | Heat exchanger |
| MPC | Model predictive control |
| P2H | Power-to-hydrogen |
| PCI | Power converter interface |
| PF | Power factor |
| TR | Thyristor-based rectifier |

### B. Indicators

| | |
|---|---|
| $b$ | Indicator of parallel stacks |
| $h, \tau$ | Temporal indicator in the hourly timescale |

### C. Variables of the Control Framework

| | | |
|---|---|---|
| $I$ | Current of the stack | A |
| $T$ | Operating temperature of the stack | ºC |
| $\delta$ | On/off state of the stack | 0-1 |
| $m^{H_2}$ | Hydrogen production rate of the stack | kg/h |
| $P$ | Power of a single P2H load | W |
| $P^{cool}$ | Cooling power | W |
| $HTO$ | Volume fraction of hydrogen to oxygen | % |
| $n^{out}$ | Volume fraction of output hydrogen impurity | % |
| $P^{P2H}$ | Power of a multi-stack P2H load | W |
| $PF$ | Power factor of a multi-stack P2H load | p.u. |

### D. Parameters

| | | |
|---|---|---|
| $n^{in}$ | Volume fraction of the input hydrogen impurity | 0.75%/h [19] |
| $HTO^{max}$ | Upper limit of HTO | 2% [19] |
| $\mathcal{F}$ | Faraday constant | 96485 C/mol |
| $p^{sep}$ | Pressure of the separator | 30 bar |
| $\mathcal{R}$ | Ideal gas constant | 8.314 J/(molK) |
| $T^{sep}$ | Temperature of the separator | 343K |
| $V^{sep}$ | Volume of the separator | 1 m³ |
| $T^{amb}$ | Ambient temperature | 25ºC |
| $U^{tn}$ | Thermo-neutral voltage | 148V |
| $R^h$ | Equivalent thermal resistance | 0.004 ºC/W [29] |
| $C^h$ | Equivalent thermal capacity | 7000 Wh/ ºC [29] |
| $N_t$ | Prediction horizon | 4h |
| $S$ | Intra-hour intervals | 4 |
| $N_B$ | Number of parallel stacks | 2/10 |
| $PF_{min}$ | The lower limit of PF | 0.9 |

## I. INTRODUCTION

Hydrogen is one of the key pillars of decarbonizing the global energy system [1]. According to the International Energy Agency (IEA) report, under the Net-zero Emissions Scenario, 80 Mt of electrolytic hydrogen production with 850 GW of power-to-hydrogen (P2H) capacity is required worldwide by 2030 [2]. In general, large-scale P2H loads consist of multi-stacks to expand the total capacity over the MW scale. For instance, Baofeng Energy's existing 100 MW P2H load is constituted by ten 5 MW single P2H loads [3]. At the same time, large-scale P2H loads are regarded as vast potential flexible resources for power systems to provide auxiliary services like peak-shaving [4]. However, the power factor (PF) is a critical index that arouses industrial attention to limit the continuous operation of P2H loads when integrating into the power grid. Under volatile conditions, the required reactive power compensation is also time-varying, which is hard to follow by switching capacitors with limited switching frequency [5]. In addition, discrete compensation may be undercompensated or overcompensated. Therefore, this paper picks a new point to solve this issue by improving control strategy by revealing the PF characteristics under volatile working conditions and involving PF considerations in the whole control framework. In the following content, the research object and issue are first described, followed by the research gaps and the main contributions.

*A. Multi-stack Power-to-Hydrogen Load*

Fig. 1 shows the research object of this paper. The grid-side parameters presented in a green dashed box are from a real constructing project. Via a 35kV/10kV transformer, reactive power compensation equipment is installed on the 10 kV common bus, paralleled with a series of single P2H loads including electrolyzers and key auxiliary facilities. In P2H loads, the core electrochemistry reaction that split water into hydrogen and oxygen with electricity injection happens in alkaline electrolyzers (AEL). Power converter interfaces (PCI) are responsible for the current conversion from alternate current (AC) to direct current (DC), and then deliver a controlled DC to the following AEL. Heat exchangers (HX) are responsible to deliver a controlled working temperature to AEL. Finally, the produced gas and lye mixture is separated in the separator.

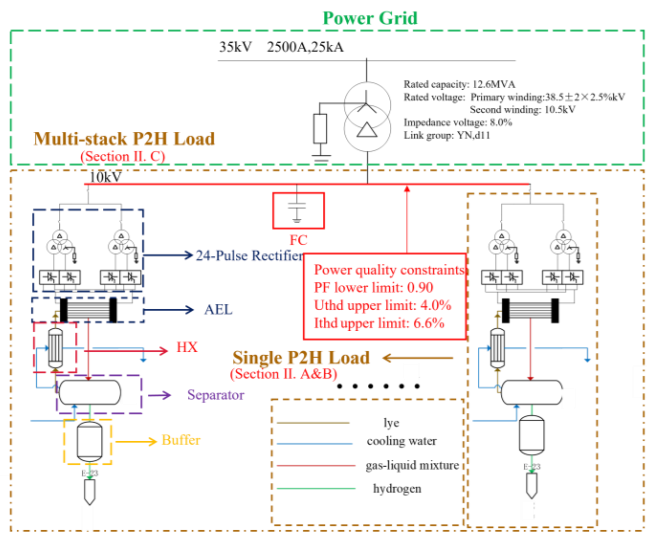

Fig. 1 The research object of a multi-stack P2H load

Electrolyzers are typically large current power loads, i.e., 200V with 5000A for a 1MW stack. For large current applications, the thyristor-based rectifier (TR) is a major group of PCI that is commonly adopted by industries [9]-[12]. TR group has priority in cost-effectiveness, however, [10], [11] figure a significant issue of TR, i.e., low power factor (PF) at the low load level on the AC side. The increment of the pulse numbers by applying phase-shifting rectifier transformers and adopting the cophase counter parallel connection [12] is an effective solution for improving the aforementioned performance. However, whether and how the PF can be accepted by the grid code is still an unknown question.

*B. Key Points of the Controller to Enable the Participation in Volatile Peak Shaving Conditions*

The integration of P2H loads into the power grid and the participation in services require balancing different operational targets and constraints, especially for the PF constraints evaluated by the power grid.

**The flexibility target**. From the perspective of the power grid, the core role of the P2H load is as a dispatchable power load. The P2H load works at variable load conditions to follow the instruction and provide flexibility to enable services like peak-shaving [4] and frequency regulation [13].

**The power factor constraint**. To ensure the normal operation of power systems, PF regulations are formulated by the operator [14]-[17]. The document [18] released by the Ministry of Water Resources and Electric Power of China stipulates that PF of power loads below 0.9 will be penalized for electricity prices.

**The production target**. From the perspective of the hydrogen operator, the core control target is to maximize hydrogen production to pursue benefits.

**The security constraint**. There are multiple security constraints in the operation of P2H loads, including feasible intervals of the voltage, current, and temperature limited by the electrodes and diaphragms in the stack [19]; on the other hand, the accumulation of hydrogen impurity in oxygen in the separator is a critical security constraint for P2H loads [20].

*C. Literature Review*

The majority of the existing research on the control strategy of multi-stack P2H loads focuses on production optimization under volatile load conditions. Guilbert et al. [21] study four-stack electrolyzers connected in parallel architecture and proposes a piecewise power-sharing strategy following wind generation. Xing et al. [22] propose a management strategy based on the production-curve model of five modules to optimize the load allocation and the switching arrangements. Flamm et al. [23] present a model-based optimal control problem to meet a given hydrogen demand in a cost-effective manner by optimizing the power dispatch of four parallel stacks. Uchman et al. [24] propose a decision method to determine the most efficient plant load distribution in three hydrogen generators.

On the physical level, in the above studies, the PF and security constraints under variable load conditions are less considered. However, different working conditions of the P2H load, i.e., determined by operating current and temperature [25], result in different PF characteristics [6] and hydrogen impurity harmful to security [23], [26]. There are risks in P2H loads to work in PF-constrained or security-constrained infeasible regions under traditional production-oriented control strategy. However, for a multi-stack P2H load, the PF characteristics under variable load conditions still lack research.

From the perspective of mathematical formation, the consideration of PF and security constraints extends the control problem from single-timescale to multi-timescale. The traditional production-oriented power distribution control follows real-time power instructions. However, the assessment of PF and the accumulation of hydrogen impurity are at least at the hourly level [18], which proposes the requirements for multi-timescale control methodology.

To fill the above gaps in the physical modeling and control methodology, this paper proposes a multi-stack P2H load control framework for the PF-constrained integration in volatile peak shaving conditions. The main contributions are two-fold:

1) The analytical models of active and reactive power of P2H loads under the inherent control variables of current and temperature are first proposed, which reveals the impact mechanism of the volatile conditions on PF characteristics. On this basis, the contrary monotonicity of PF and production efficiency reveals an intrinsic control tradeoff between the PF constraint and the production target.

2) A PF-constrained multi-timescale control framework is proposed and verified to guarantee the control accuracy and tractability, including a model-based hour-ahead robust model predictive controller (MPC) and a rule-based real-time increment correction algorithm. The significance of the extended PF and security consideration in improving the flexibility target is verified. Furthermore, the typical operating modes at the cluster level are captured for the reference of industrial applications.

The remainder of the paper is organized as follows: Section II analyzes power characteristics of the P2H load in volatile conditions. Section III discusses the tradeoff between the PF constraint and the production target. Section IV proposes the PF-constrained multi-timescale control framework. In Section IV, case studies are conducted to compare the proposed control framework and traditional control strategy, and propose the cluster operation modes. The summary and conclusions follow in Section VI.

## II. THE POWER CHARACTERISTICS OF THE P2H LOAD IN VOLATILE CONDITIONS

As aforementioned in Section I.A shown in Fig. 1, the whole P2H load includes the complex interaction of electricity, mass, and heat. Electrolyzers are core components in P2H loads to convey the influence of non-electrical parameters like temperatures on electrical characteristics like voltages and power via electrochemical reactions. Therefore, this section first builds the bridge of the UI Characteristic on the DC side with the internal operation parameters of P2H. Then, the UI Characteristic on the AC side is modeled as functions of the DC-side UI. On this basis, the active and reactive power characteristics can be captured by the AC-side UI. Finally, the influence of the internal parameters of P2H on its electrical performance can be analyzed.

### A. The UI Characteristic on the DC side
*a. The Relationship of $U$ with $I$ and $T$*

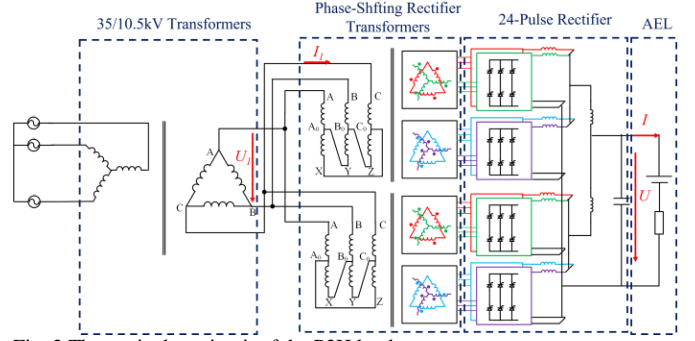

Fig. 2 The equivalent circuit of the P2H load

Fig. 2 describes the electrical part of the P2H load, where PCI connects the AC side and the DC side. On the DC side, the electrolysis stack can be equivalent to a series connection of a electromotive force and a resistance, which satisfies the electrochemical equation expressed as (1). The external voltage $U$ consists of two main parts [19], [25]: $U_{rev}$ is the minimum voltage required to decompose water that is related to Gibbs energy; $IR$ is the overpotential, $R$ is the equivalent resistance considering the ohmic losses on the electrolyte, the formation of the gas bubbles on the electrode surfaces, the diaphragm, and the activation losses on the electrodes caused by the electrode kinetics. The electrochemical equation determines that the voltage is determined by two significant control variables, i.e., operating current $I$ and temperature $T$.

$$\begin{aligned}
U(I,T) &= U_{rev}(T) + IR(I,T) \\
&= 9.84 \times 10^{-7}(T+273)^2 - 0.0015T \\
&+ \log\left(\begin{array}{l} \exp\left(0.5 - \log\left(\exp\left(5 - \dfrac{1.7 \times 10^3}{t+233}\right)\right)\right) \\ \exp\left(5 - \dfrac{1.7 \times 10^3}{t+233}\right) \\ \left(\left(29 - \exp\left(\log\left(\exp\left(5 - \dfrac{1.7 \times 10^3}{t+233}\right)\right)\right) - 0.5\right)^{1.5}\right) \end{array}\right) \\
&\quad \left(4.3 \times 10^{-5}T + 0.01\right) + \log(T+273)\left(9.5 \times 10^{-5}T + 0.02\right) \\
&+ \dfrac{0.06I}{\left(1 - 1.5 \times 10^{-3}\left(\dfrac{I}{300}\right)^{0.3}\right)^{1.5}} \\
&\quad \left(0.04T + \dfrac{1.6 \times 10^3}{T+273} - 2 \times 10^{-5}(T+273)^2 - 4\right) \\
&+ \dfrac{\log(317I - 10)(19T + 5 \times 10^3)}{183T + 6 \times 10^4} \\
&+ \log(317I - 10)\left(\dfrac{19T + 5 \times 10^3}{183T + 6 \times 10^4}\right) + 1
\end{aligned} \quad (1)$$

*b. The Intervals of $I$ and $T$*

In P2H loads, PCI controls $I$ and HX controls $T$. The electrode material puts caps on the upper limits of $I$ and $T$.

For instance, for a 1 MW P2H load, $I \in [0,5]$ kA and $T \in [30,80]$ °C. Furthermore, different operation statuses correspond to different intervals of $I$ and $T$, as shown in Fig. 3(a). On Normal status, $I$ is adjustable in the feasible interval and $T$ maintains a high level to improve efficiency. When power input is temporarily cut, $I=0$ corresponds to Standby status. However, if there is a long-duration gap of power input, the whole P2H load needs to shut down to stay at an Off status with $\delta = 0$. The steady $T$ on Off status is consistent to the ambient temperature. As the P2H load starts up, $T$ is gradually rising due to the exothermic electrochemical reaction.

Considering the all above statuses, the general intervals of $I$ and $T$ related to the operating status are expressed as (2)-(3).

$$I_t \leq I_{\max} \delta_t \quad (2)$$

$$T_{\min} \leq T_t \leq T_{\max} \quad (3)$$

Therefore, under the general intervals of $I$ and $T$, the external voltage $U$ and the equivalent resistance $R$ are determined and show the feasible regions in Fig. 3(b) and (c). For a 1 MW P2H load, the detailed expression of $U$ is expressed as (1) [29], which reflects that $U$ is positively correlated to $I$ and negatively correlated to $T$.

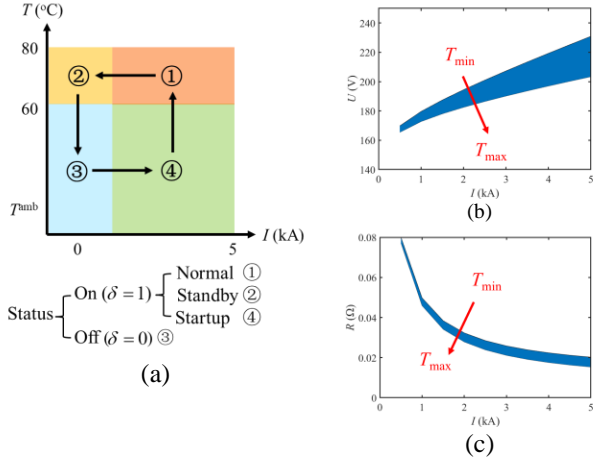

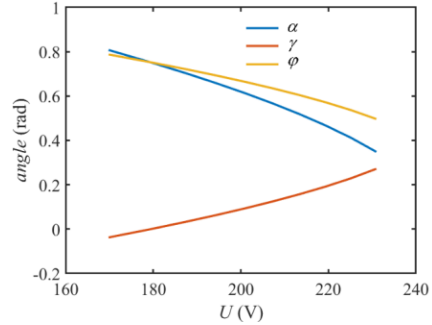

Fig. 3 (a) Different operation statuses of P2H loads and the corresponding intervals of $I$ and $T$; (b) and (c) The external voltage and the equivalent resistance determined by $I$ and $T$

### B. The UI Characteristic on the AC side

We design a 24-TR topology based on the cophase counter parallel connection of two 24-pulse TR, as shown in Fig. 2. The following model parameters are obtained by the simulation on the established Simulink model shown in [27].

*a. The Power Factor Angle*

For a 24-pulse topology, the ideal relationship of the effective value of voltage on the AC-side $U_1$ and $U$ satisfies the following equation [27]:

$$U = \frac{2.44}{K} U_1 \cos \alpha \Leftrightarrow \alpha(U) = \arccos\left(4.07 \times 10^{-3} U\right) \quad (4)$$

where $K=104$ is the turn ratio of the transformer, $U_1=10.465$ kV is assumed as a constant determined by the grid, $\alpha$ is the firing angle.

However, considering the realistic influence of the leakage reactance of the transformer, $\alpha$ is required to be corrected with a commutation overlap angle $\gamma$. $\gamma$ is negatively correlated to $\alpha$ and can be fitted by (5). Finally, the real power factor angle $\varphi$ satisfies the correction equation in (6).

$$\gamma(\alpha) = -0.6738\alpha + 0.5065 \quad (5)$$

$$\cos \varphi = \frac{\cos \alpha + \cos(\alpha + \gamma)}{2} \quad (6)$$

According to the interval of $U$ determined by (1)-(3), the intervals of the aforementioned three angles are shown in Fig. 4. $\varphi$ reflects a negative correlation to $U$ due to the monotonicity of the arccos function.

Fig. 4 The intervals of $\alpha$, $\gamma$, and $\varphi$

*b. The Relationship of the AC-side UI with the DC-side UI*

After the correction of $\varphi$, the relationship of the AC-side UI with the DC-side UI can be determined by (7)-(8). (8) is established based on the power conservation law. The left side is the sum active power expressed with $U_1$, the fundamental current component $I_1$, and $\varphi$. The right side is expressed as the addition of power consumption on the electrolysis stack expressed as $UI$ and the power loss on the PCI, which mainly includes the fixed transformer iron loss that is unrelated to $I$, the device switching loss that is approximately proportional to $I$, and the device on-state loss, transformer copper loss, capacitor and inductor equivalent resistance loss that are approximately proportional to the second order of $I$ [30]. Therefore, the power loss on PCI is fitted by a quadratic polynomial equation [31]:

$$U = \frac{2.44}{K} U_1 \cos \varphi \quad (7)$$

$$\sqrt{3} U_1 I_1 \cos \varphi = UI + aI^2 + bI + c \quad (8)$$

where $a$, $b$, and $c$ are coefficients related to parameters of devices and transistors.

Except for the fundamental current component $I_1$, the 24-TR topology causes the current distortion during the AC/DC conversion. The conversion of the total distortion current component $I_d$ to $I_1$ is via a fundamental component factor $v$ ($v=0.9971$ for 24-pulse topology). Finally, based on (7)-(8), the current effective value on the AC-side can be expressed.

$$I_d = \frac{\sqrt{1-v^2}}{v} I_1$$
$$= 1.033 \times 10^{-3} \left( I + \frac{5.4 \times 10^{-4} I^2 + 1.353 I + 91940}{U} \right) \quad (9)$$

*C. The Power Characteristics*

The electrochemical equation in (1) connects $U$ with $I$ and $T$, and (7)-(9) connect $U_1$, $I_1$, $I_d$ with $U$, $I$. Finally, in this subsection, active and reactive power $P,Q$ can be expressed by $U$, $I$ that are intrinsically determined by $I$ and $T$.

*a. The Active Power Characteristics*

According to (8), the sum active power of P2H loads is mainly from AELs and PCIs, expressed as (10).

$$P(I,T)=U(I,T)I+5.4\times10^{-4}I^2+1.353I+91940 \quad (10)$$

(10) obviously reflects the positive correlation of $P(I,T)$ with $I$ and the negative correlation with $T$ according to the monotonicity of (1). Following the parameters in [19], a typical $P-I-T$ region of a 1MW alkaline stack that shown in Fig. 5 ranges from 0.24MW to 1.21MW.

*b. The Reactive Power Characteristics*

The sum reactive power consists of three parts: the phase shift component $Q_s$, the distortion component $Q_d$, and the constant compensation component $Q_c$, expressed in (11) [32].

$$Q(I,T)=\sqrt{Q_s^2+Q_d^2}-Q_c \quad (11)$$

$$\begin{aligned}Q_s(I,T) &= \sqrt{3}U_1 I_1 \sin\varphi \\ &= 245.525\left(I+\frac{5.4\times10^{-4}I^2+1.353I+91940}{U(I,T)}\right)\times\ldots \\ &\quad \sin\left(0.6339\arccos\left(4.07\times10^{-3}U(I,T)\right)+0.2756\right)\end{aligned} \quad (12)$$

$$\begin{aligned}Q_d(I,T) &= \sqrt{3}U_1 I_d \\ &= 18.716\left(I+\frac{5.4\times10^{-4}I^2+1.353I+91940}{U(I,T)}\right)\end{aligned} \quad (13)$$

Based on (6)-(9), the expressions of $Q_s$ and $Q_d$ can be expressed as (12)-(13). $Q(I,T)$ reflects the positive correlation with $I$ and $T$ according to monotonicity of (1). Following the parameters in [19], a typical $Q-I-T$ region of a 1MW alkaline stack that shown in Fig. 6 ranges from 0.08MVar to o,52MVar.

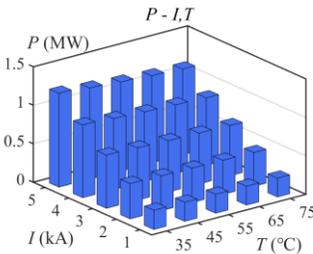
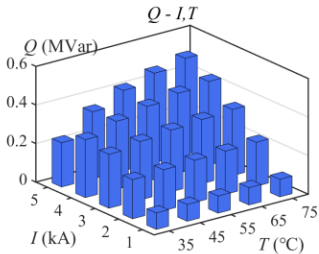

Fig. 5 Active power characteristics     Fig. 6 Reactive power characteristics

### III. THE TRADEOFF BETWEEN THE POWER FACTOR CONSTRAINT AND THE PRODUCTION TARGET

This section captures the PF characteristic of a multi-stack P2H load in volatile power conditions and analyzes its impact on the production-oriented optimal allocation rule. First, following the power characteristics obtained in Section II, the monotonicity of the PF characteristic is analyzed and compared with the efficiency characteristic. Then, the PF characteristic of a two-stack P2H load is presented to show the inherent PF-constrained control regions. Finally, PF-constrained production-oriented optimal allocation rule is discussed.

*A. The Contrary Monotonicity of PF and Efficiency*

*a. PF*

According to the research object shown in Fig. 1, PF on the 10 kV side is evaluated, which is determined by the active power and reactive power that can be expressed in (14) [33]. It reveals that PF is related to the operation conditions $I,T$ of stacks.

$$PF(I,T)=\frac{P(I,T)}{\sqrt{(P(I,T))^2+(Q(I,T))^2}} \quad (14)$$

*b. Production Efficiency*

The quantity of hydrogen production $m^{H_2}$ can be determined by (15)-(16). $\eta^I$ is the current efficiency related to $I$ [29]. On this basis, the hydrogen production efficiency $\eta(I,T)$ can be calculated by (17), which is also related to the operation conditions $I,T$ of stacks. LHV =33.33kWh/kg is the lower heating value of hydrogen.

$$m^{H_2}(I)=360\eta^I I/\mathcal{F} \quad (15)$$

$$\eta^I=0.96I^2/(250000+I^2) \quad (16)$$

$$\eta(I,T)=\frac{m^{H_2}(I)\text{LHV}}{P(I,T)} \quad (17)$$

Here, we analyze the monotonicity of $PF(I,T)$ and $\eta(I,T)$ on the feasible domain of $I,T$. Fig. 7-8 show the values of $PF(I,T)$ and $\eta(I,T)$. Following the models in Section II, $PF(I,T)$ presents a "valley" characteristic, i.e., a first-decrease-then-increase tendency with $I$ increasing, and a monotonically decreasing trend with $T$ increasing. In contrast, $\eta(I,T)$ presents a "hill" characteristic, i.e., a first-increase-then-decrease tendency with $I$ increasing, and a monotonically increasing trend with $T$ increasing. The contrary monotonicity of $PF$ and $\eta$ reveals an intrinsic tradeoff between the PF constraint and the production target.

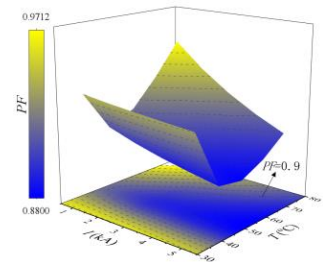
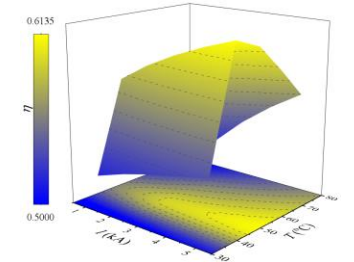

Fig. 7 Valley characteristic of PF     Fig. 8 Hill characteristic of efficiency

Furthermore, there is PF-constrained infeasible region where PF<0.9, as shown in Fig. 7, which requires the attention in extended multi-stack P2H loads.

*B. PF-Constrained Production-oriented Allocation Rule*

*a. PF-constrained Domain*

Fig. 9 shows the PF characteristics of a two-stack P2H load via 24-TR. Fig. 9 plots the $I_{stack1}$ and $T_{stack1}$ of the first stack on the horizontal axis against $I_{stack2}$ and $T_{stack2}$ of the second stack on the vertical axis. The color bar is listed in the righthand, and different color mapped in Fig. 9 represents various PF values. The operation points that PF lower than 0.9 are circled with red dotted lines. According to the "valley" characteristic shown in Fig. 7, for a two-stack P2H load, there are PF-constrained infeasible regions represented by red dashed boxes in Fig. 9. Moreover, according to the monotonicity of PF with $T$, it represents that for 24-TR topology, the higher $T$, the larger PF-constrained infeasible region.

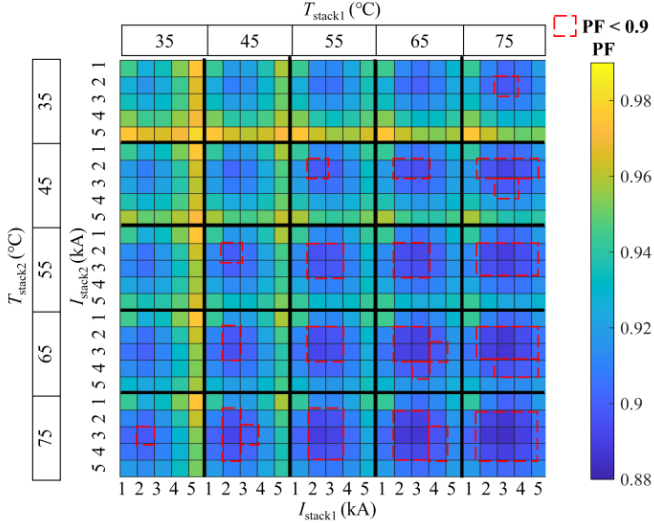

Fig. 9 PF characteristics of a two-stack P2H load

*b. PF-constrained Optimal Allocation Rule*

For a two-stack P2H load, once the sum load power $P$ is given, the PF-constrained optimal control strategy of $I_1, I_2, T_1, T_2$ to pursue the maximum production can be obtained by solving the optimization programming in (18).

$$\max_{I_1,I_2,T_1,T_2 \in \mathcal{D}} m^{H_2} = m_1^{H_2}(I_1) + m_2^{H_2}(I_2)$$
$$s.t. \begin{cases} P_1(I_1,T_1) + P_2(I_2,T_2) = P \\ \mathcal{D} = \{I_1, I_2, T_1, T_2 \mid PF(I_1,I_2,T_1,T_2) \geq 0.9\} \end{cases} \quad (18)$$

According to our previous research [22], [25], the production-oriented optimal control without PF consideration obeys the "equal-split" rule, i.e., stacks work on the same $I$ points to pursue the maximum efficiency, which is determined by the concave characteristic of the efficiency profile shown in Fig. 8. And the higher $T$, the lower energy consumption and the higher efficiency. However, the PF constraint shrinks the feasible domain. Here, the optimal allocation rule under three status portfolios is analyzed, the results are shown in Fig. 10.

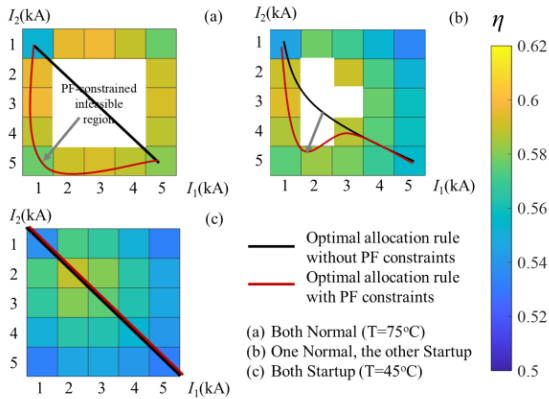

Fig. 10 PF-constrained optimal allocation rule

When both stacks work on Normal statuses, i.e., $T_1$ and $T_2$ maintain at the high-temperature level, there is a large PF-constrained infeasible region according to Fig. 9. The original optimal "equal-split" rule of $I_1 = I_2$ along the diagonal line needs to twist, especially at the medium load level, as shown in Fig. 10(a). Finally, the current unequal-split rule is accepted. In contrast, if both stacks work on Startup statuses at a low-temperature level, as shown in Fig. 10(c), the PF constraint is slack to ensure the current "equal-split" optimal allocation rule. The result of the circumstance (b) is within that of (a) and (c). The different working temperature pulls the original allocation solution to the higher-temperature side (stack2), while PF constraints further intensify this imbalance, as shown in Fig. 10(b).

In summary, this section verifies the necessity of the consideration of the PF constraint and its influence on the optimal allocation rule. In the next section, PF constraints are embedded into the whole control framework of a multi-stack P2H load.

## IV. POWER FACTOR-CONSTRAINED MULTI-TIMESCALE CONTROL FRAMEWORK OF A MULTI-STACK P2H LOAD

This section proposes the PF-constrained multi-timescale control framework. The controller needs to achieve the real-time current distribution following the peak-shaving instructions from the grid operator and ensure the security constraints. However, the startup and shutdown actions and the assessment of PF are at least on an hourly timescale. Therefore, a two-level control framework is proposed, which consists of an hour-ahead robust model predictive controller (MPC) and a real-time increment correction algorithm, as shown in Fig. 11.

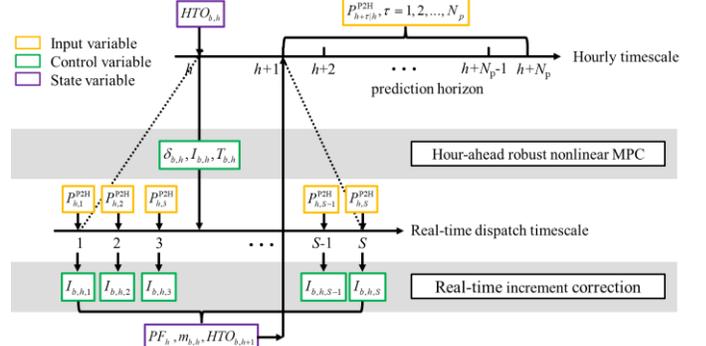

Fig. 11 Illustration of the multi-timescale controller

At period $h$, hour-ahead robust MPC is responsible for providing a decision on $\delta_{b,h}$, $T_{b,h}$ and a baseline $I_{b,h}$ during $[h, h+1)$ based on the power instruction $P_{h+\tau|h}^{P2H}, \tau = 1, 2, ..., N_p$ and $HTO_{b,h}$ of the last period. And at sub-period $s$ in $[h, h+1)$, the increment correction of $I_{b,h,s}$ based on $I_{b,h}$ can be determined with the real-time power instruction input of $P_{h,s}^{P2H}$. At the end of period $h$, the actual state variable $PF_h$, $m_{b,h}$, and $HTO_{b,h+1}$ are exactly evaluated and back to the hour-ahead MPC for the decision at period $h+1$. The two-level current distribution controller is depicted in detail as follows.

### A. Hour-ahead Robust MPC

*a. The Original Nonlinear MPC*

At each control period $h$, the hour-ahead robust MPC is expressed as the following optimization programming. To address the uncertainty of the real-time power instruction deviation to ensure the robustness of the PF constraint, a standard robust optimization model is set up to make the optimal control strategies against the "worst" case. Here, $P_{h+\tau|h}^{P2H}$ is the power instruction value of the period $h+\tau$ from the dispatch center at the period $h$. $\Delta P_{h+\tau|h}^{P2H}$ is the possible real-time power instruction deviation within the period $h$. $\tau = 0,1,...,N_p -1$, $Np$ denotes the prediction horizon. $HTO_{b,h+\tau+1|h}$ is the state variable, $I_{b,h+\tau|h}, T_{b,h+\tau|h}, \delta_{b,h+\tau|h}$ are control variables.

$$\max_{I,T,\delta} \max_{\Delta P_{h+\tau|h}^{P2H}} J = \sum_{\tau=0}^{N_p-1} \sum_{b=1}^{N_B} m_{b,h+\tau|h}^{H_2}(I) \quad (19)$$

for $\forall \tau = 0,1,...,N_p -1$:

$$\sum_{b=1}^{N_B} P_{b,h+\tau|h}(I,T) = P_{h+\tau|h}^{P2H} + \Delta P_{h+\tau|h}^{P2H} \quad (20)$$

$$|\Delta P_{h+\tau|h}^{P2H}| \le \alpha P_{h+\tau|h}^{P2H}$$

$$\sum_{\tau=0}^{N_p-1} PF_{h+\tau|h}(I,T) \ge N_p PF_{\min} \quad (21)$$

$$(2)-(3) \quad (22)$$

The objective of the MPC controller in (19) is to maximize hydrogen production determined by (15)-(16). The constraints include: (20) represents the load power division of the power instruction between stacks considering $\Delta P_{h+\tau|h}^{P2H}$ bounded by the coefficient $\alpha$. (21) represents that the average PF index $PF$ should be in the levels limited by the grid code. (22) represents the inner secure operation constraints of P2H loads that are the same as (2)-(3).

Moreover, according to our previous research [20], [25], $I$ and $T$ should obey one-order constraints due to the inertia of the accumulation of gas and heat. Limited by the cross-diffusion of product gasses that can result in flammable gas mixtures, $HTO$ in the separator should be in the safe interval. (23) describes the one-order dynamic accumulation process of $HTO$ as the function of $I$. The volume fraction of the output hydrogen impurity $n^{out}$ in (24) is related to the oxygen production rate determined by $I$ [20].

$$HTO_{b,h+\tau+1|h} = HTO_{b,h+\tau|h} + n_{b,h+\tau|h}^{in} - n_{b,h+\tau|h}^{out} \quad (23)$$

$$n_{b,h+\tau|h}^{out} = HTO_{b,h+\tau|h} \frac{\mathcal{R}T^{sep}}{p^{sep}V^{sep}} \frac{\eta_{b,h+\tau|h}^I I_{b,h+\tau|h}}{4\mathcal{F}} \quad (24)$$

$$HTO_{b,h+\tau|h} \le HTO_{\max} \quad (25)$$

For the constraint on $T$, (26) describes the one-order dynamic equation of $T$, where the cooling power $P_t^{cool}$ controls the change of $T$ [25].

$$T_{b,h+\tau|h} = T_{b,h+\tau-1|h} + \frac{1}{C^h}\left(P_{b,h+\tau|h}^{AEL} - U^{tn}I_{b,h+\tau|h} - \frac{1}{R^h}\left(T_{b,h+\tau|h} - T^{amb}\right) - P_{b,h+\tau|h}^{cool}\right) \quad (26)$$

In summary, (19)-(26) describe the original hour-ahead robust MPC model.

*b. The Piecewise Reformulation*

$m_b^{H_2}$ and $P_b$ are nonlinear functions of $I_b$ and $T_b$, which can be dealt with the piecewise linearization method and further expressed as:

$$m_b^{H_2}(I) = \sum_{k=1}^{N_I}\left(A_{b,k}I_{b,k} + B_{b,k}\sigma_{b,k}\right) \quad (27)$$

$$P_b(I,T) = \sum_{k=1}^{N_I}\sum_{l=1}^{N_T}\left(C_{b,l,k}I_{b,k}\lambda_{b,l} + D_{b,l,k}T_{b,l}\sigma_{b,k} + E_{b,l,k}\sigma_{b,k}\lambda_{b,l}\right) \quad (28)$$

where $\sigma_{b,k}=1$ indicates that $I_b$ belongs to the $k^{th}$ segment $I_{b,k}$, and $\lambda_{b,l}=1$ indicates that $T_b$ belongs to the $l^{th}$ segment $T_{b,l}$. $A_{b,k}, B_{b,k}, C_{b,l,k}, D_{b,l,k}, E_{b,l,k}$ are coefficients.

Furthermore, $PF(I,T)$ of a multi-stack P2H load can be fitted by the polynomial formation based on the adaptive sparse Smolyak grid sampling method [34], [35] sampled from the established Simulink simulation model. Here, to make sure the compatibility of the PF expression that suitable for the controller, cross terms related to different stacks are not reserved to improve solvability. The fitting error of PF of a 10-stack P2H load is shown in Fig. 12. Therefore, cubic polynomial piecewise formation in (29) with the discretization of $I,T$ is adopted with the fitting error below 2%.

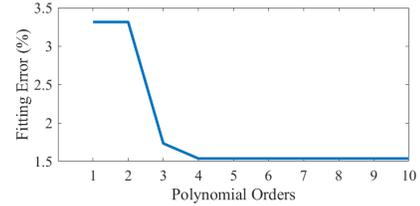

Fig. 12 The relationship of the fitting error with polynomial orders

$$PF(I,T) = d + \sum_{b=1}^{N_B}\left(\begin{array}{l} e\sum_{k=1}^{N_I}I_{b,k} + f\sum_{l=1}^{N_T}T_{b,l} + g\sum_{k=1}^{N_I}(I^k)^2\sigma_{b,k} + \\ h\sum_{k=1}^{N_I}\sum_{l=1}^{N_T}I^kT^l\sigma_{b,k}\lambda_{b,l} + z\sum_{k=1}^{N_I}(I^k)^3\sigma_{b,k} \end{array}\right) \quad (29)$$

where $d,e,f,g,h,z$ are the fitting coefficients.

In this way, the original hour-ahead MPC is reformulated in the form of MILP. The proposed bi-level robust programming in can take dual for the inner "max" linear programming model according to strong duality [36], which yields a single level "max" model.

### B. Real-time Increment Correction Algorithm

Once $I_{b,h}$ and $\delta_{b,h}$ are determined, during the time interval $[h, h+1)$, the control variable $I_{b,h,s}$ can be expressed as the addition of $I_{b,h}$ and an increment correction $\Delta I_{b,h,s}$, while the latter should be calculated following the real-time power instruction. The real-time optimal increment distribution pursues the maximum energy conversion efficiency that is closely related to two indexes: $\dfrac{\mathrm{d}m_b^{H_2}(I_{b,h})}{\mathrm{d}I_b}$ represents the increment rate of hydrogen production with $\Delta I_{b,h,s}$ on $I_{b,h}$, and $\dfrac{\mathrm{d}P_b(I_{b,h})}{\mathrm{d}I_b}$ is the increment rate of power consumption with $\Delta I_{b,h,s}$ on $I_{b,h}$. Therefore, the priority of the distribution of the current increment is determined by the index of $rank_{b,h} = \dfrac{\mathrm{d}m_b^{H_2}(I_{b,h})}{\mathrm{d}I_b} / \dfrac{\mathrm{d}P_b(I_{b,h})}{\mathrm{d}I_b}$. Specifically, when $\Delta P_{h,s}^{P2H} > 0$, the priority order of the current increment distribution is positively correlated with $rank_{b,h}$, and vice versa. However, the increment distribution $\Delta I_{b,h,s}$ should also ensure that $HTO_{b,h,s} \leq HTO^{max}$ always holds, which is calculated by:

$$\Delta HTO_{b,h,s+1} = \Delta HTO_{b,h,s} + \dfrac{f_b(HTO_{b,h}, I_{b,h})}{S} + \dfrac{\partial f_b(HTO_{b,h}, I_{b,h})}{\partial HTO_b}\dfrac{\Delta HTO_{b,h,s}}{S} + \dfrac{\partial f_b(HTO_{b,h}, I_{b,h})}{\partial I_b}\dfrac{\Delta I_{b,h,s}}{S} \quad (30)$$

$$HTO_{b,h,s} = HTO_{b,h} + \Delta HTO_{b,h,s} \quad (31)$$

where the detailed expression of $f_b(HTO_{b,h}, I_{b,h}) = n_{b,h}^{in} - n_{b,h}^{out}$ is given by (23)-(24).

Therefore, the real-time increment calculation algorithm of $I_{b,h,s}$ is shown in TABLE I.

TABLE I
THE REAL-TIME INCREMENT CALCULATION ALGORITHM OF $I_{b,h,s}$

| | |
|---|---|
| 1: | **input** $P_h^{P2H}$, $\Delta P_h^{P2H}$, $I_{b,h}$, $HTO_{b,h}$, $rank_{b,h}$ |
| 2: | $s=1$ |
| 3: | **while** $s<S+1$ **do** |
| 4: | **input** $P_{h,s}^{P2H}$ |
| 5: | $\Delta P_{h,s}^{P2H} = P_{h,s}^{P2H} - P_h^{P2H} - \Delta P_h^{P2H}$ |
| 6: | **if** $\Delta P_{h,s}^{P2H} \geq 0$ { $\Delta P_{h,s}^{P2H} < 0$ } **then** |
| 7: | **for** $b$ in the ascending{descending} order of $rank_{b,h}$ |
| 8: | $\Delta I_{b,h,s} = \left(\Delta P_{h,s}^{P2H} - \sum_{l=1}^{b-1}\left(\dfrac{\mathrm{d}P_l(I_{l,h})}{\mathrm{d}I_l}\Delta I_{l,h,s}\right)\right)/\dfrac{\mathrm{d}P_b(I_{b,h})}{\mathrm{d}I_b}$ |
| 9: | calculate $\Delta HTO_{b,h,s}$ and $HTO_{b,h,s}$ based on (30)-(31) |
| 10: | **if** $HTO_{b,h,s} > HTO^{max}$ **then** |
| 11: | $\Delta I_{b,h,s} = S\left(\Delta HTO_{b,h,s} - \dfrac{f_b(HTO_{b,h}, I_{b,h})}{S} - \left(1+\dfrac{1}{S}\dfrac{\partial f_b(HTO_{b,h}, I_{b,h})}{\partial HTO_b}\right)\Delta HTO_{b,h,s-1}\right)/\dfrac{\partial f_b(HTO_{b,h}, I_{b,h})}{\partial I_b}$ |
| | /*real-time security constraint*/ |
| 12: | **end if** |
| 13: | $\Delta I_{b,h,s} = \min\{\Delta I_{b,h,s}, I_{max} - I_{b,h}\}$ { $\Delta I_{b,h,s} = \max\{\Delta I_{b,h,s}, -I_{b,h}\}$ } |
| 14: | /*real-time upper and lower limits*/ |
| 15: | **end** |
| 16: | **end if** |
| 17: | calculate $\Delta HTO_{b,h,s}$ and $HTO_{b,h,s}$ based on (30)-(31) |
| 18: | $I_{b,h,s} = I_{b,h} + \Delta I_{b,h,s}$ |
| 19: | **output** $I_{b,h,s}$ |
| | $s=s+1$ |
| 20: | **end** |

At the end of period $h$, the state variable $m_{b,h}$ can be determined by (15)-(16) based on the actual $I_{b,h,s}$. $HTO_{b,h+1}$ is back to the hour-ahead MPC for the decision at period $h+1$, which is determined by (32).

$$HTO_{b,h+1} = HTO_{b,h} + \Delta HTO_{b,h,S} \quad (32)$$

## V. CASE STUDIES

Case studies are organized as follows: first, the control strategy based on the proposed framework is analyzed and compared with the traditional production-oriented control strategy based on a two-stack P2H load. Moreover, the proposed control framework is applied in a multi-stack P2H load to explore the clustering characteristic and verify the necessity and effectiveness of the proposed control framework.

The hour-ahead MPC is performed on a computer with an i9-10980XE CPU and 128 GB of memory. The programming is developed using Matlab R2020b and solved by IBM ILOG CPLEX v12.10.0 with branch and bound algorithm.

### A. Case Setup
#### a. Object and Parameters
The multi-stack P2H load with parameters shown in Fig. 1 is chosen in case studies. According to the project circumstance, PF is evaluated on the 10 kV side, as shown in Fig. 1. The capacity of one stack is 1MW. The detailed values of coefficients in (27)-(29) are shown in [37].

#### b. Scenario and Indexes
A typical peak-shaving instruction is considered to be an anti-load profile, as shown in red profile $P^{P2H}$ in Fig. 13. At peak-load periods, P2H reduces the load power; and at valley-load periods, P2H increases the load power to consume the extra renewable energy. Here, $\alpha = 5\%$ is considered.

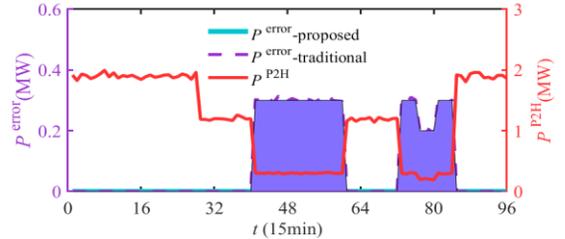

Fig. 13 $P^{P2H}$ and $P^{error}$ of a two-stack P2H load for proposed and traditional strategies respectively

The control performances are evaluated by the following three indexes.

*1)* **Flexibility**: the cumulative error in response to power instructions, expressed as $\sum_{h=1}^{24}\sum_{s=1}^{4}\left|P_{h,s}^{P2H} - \sum_{b=1}^{N_B}P_{b,h,s}\right|$.

*2)* **Power factor**: the average PF evaluated by the real-time control strategies, expressed as $\dfrac{1}{96}\sum_{h=1}^{24}\sum_{s=1}^{4}PF_{h,s}$.

*3)* **Production**: the total hydrogen production, expressed as $\sum_{h=1}^{24}\sum_{s=1}^{4}\sum_{b=1}^{N_B}m_{b,s}^{H_2}$.

### B. Comparison with the Traditional Control Strategy based on a Two-stack P2H Load

The traditional control strategy is defined as follows:
*1)* To ensure the satisfaction of the security constraint, the lower limit of $I$ is generally considered to be 40% in (2) [19].
*2)* The PF and security constraints in (21), (23)-(25), are not involved in the hour-ahead MPC, and steps 9-12, 17 in TABLE I are not executed in the real-time current correction process.

The control performances of two control strategies are concluded as TABLE II.

TABLE II

TABLE II
THE CONTROL PERFORMANCE OF TWO STRATEGIES

| Strategy | Flexibility (MW) | Power factor (p.u.) | Production (kg) |
|---|---|---|---|
| traditional | 9.19 | 0.9036 | 491.06 |
| proposed | 0 | 0.9181 | 534.90 |

**Hourly timescale control strategy**

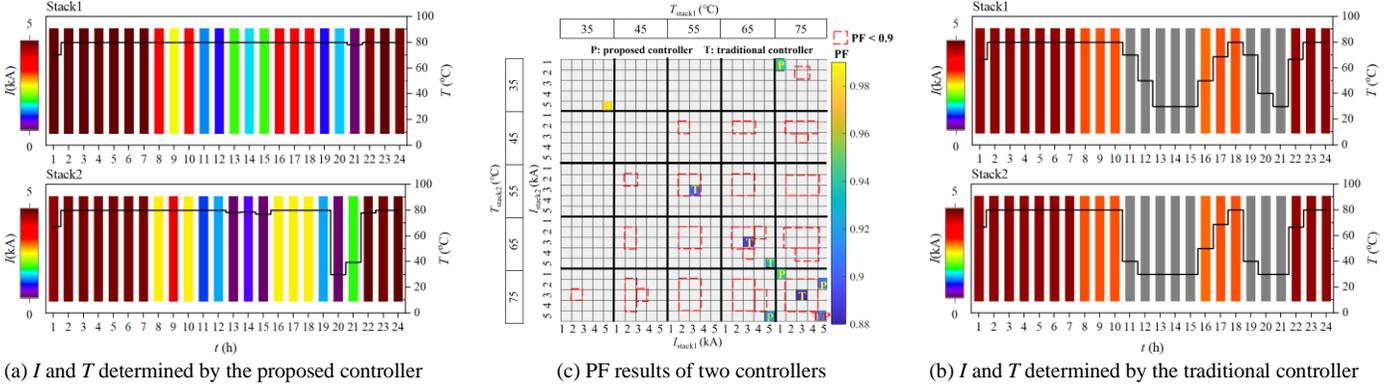

(a) $I$ and $T$ determined by the proposed controller
(c) PF results of two controllers
(b) $I$ and $T$ determined by the traditional controller

**Real-time dispatch timescale control strategy**

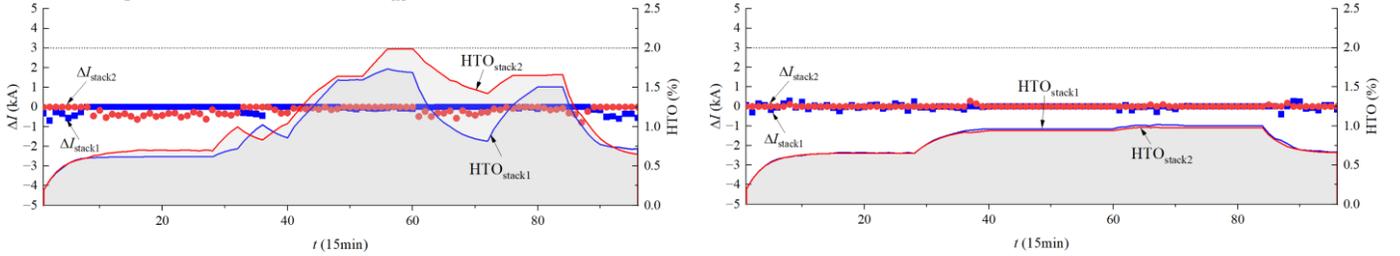

(d) $\Delta I$ and HTO results of the proposed controller
(e) $\Delta I$ and HTO results of the traditional controller

Fig. 14 Comparison of control strategies between the proposed controller and the traditional controller in a two-stack P2H load.

*a. Advantages on Power Factor*

The control strategy proposed in this paper has obvious advantages over the traditional control strategies in terms of PF, as shown in Fig. 14(c). Fig. 14(c) imitates the form of Fig. 9. Operation points related to PF results shown in Fig. 14(c) are presented in Fig. 14(a) and (b) under the proposed controller and the traditional controller, respectively. Traditional controller obeys the current equal-split law without considering PF constraints, resulting in the operation points located at the main diagonal of Fig. 14(c), especially at red dotted boxes where PF is lower than 0.9. In contrast, operation points of the proposed controller perfectly avoid the PF-constrained infeasible region. Under the same PCI and load conditions, the average PF of the proposed is 0.9163, while the traditional one is 0.9070, as shown in TABLE II.

As aforementioned PF characteristics of 24-TR topology, it reveals that the unequal division of current corresponds to a higher PF. Therefore, in response to the power instruction, an unequal-split current control strategy is adopted at the medium load level during 8~10 h and 16~18 h, as shown in Fig. 14(a). Taking the time from 16 h to 18 h as an example, the current of one stack is 4 kA, and the current of the other stack is 2.53 kA, resulting in PF to be 0.9025. In comparison, the traditional controller determines an equal-split current strategy to be 3.14 kA during this period, as shown in Fig. 14(b), accompanied with PF to be 0.8845 that is lower than the requirement of the grid code.

Meanwhile, excellent PF is obtained at the expense of reduced hydrogen production. During 8~10 h and 16~18 h, the proposed control strategy scarifies 2.91 kg hydrogen production compared to the traditional strategy. It verifies the existence of the tradeoff between PF and production.

Moreover, Fig. 14(d) shows that the majority of $\Delta I$ in the real-time correction is negative, which means that $\Delta P_{h+\tau|h}^{P2H} > 0$ is considered in the hour-ahead robust control to ensure the robustness of PF. The real-time PF according to the actual control strategy is shown in Fig. 14(c), all working points meet the requirements, which verifies the robustness of PF under the proposed control framework.

*b. Advantages on Flexibility and Production*

From Fig. 13, when $t$=40~60, 72~84 (15min), the traditional control strategy cannot fully track the power instruction at low loads, leading to a 9.19 MW daily error compared to the proposed control strategy, as shown in TABLE II.

The conservative lower limit of $I$ in traditional strategies sacrifices the actual ability of electrolyzers to follow the instruction. Fig. 14(b) shows that during $t$=40~60 (15min) and 72~84 (15min), both stacks shut down at the low-load condition under the traditional strategy. Actually, the minimum operating current is limited by the HTO security constraint. The fixed set of the lower limit of current is conservative since the highest HTO under the traditional strategy just attains 1 which is far from the upper limit of 2, as shown in Fig. 14(e). In the proposed control strategy, HTO security constraint rather than the fixed lower ratio of $I$ is considered. Therefore, in this scenario, the minimum operating current can temporarily be 0 A to stay in the standby status rather than the shutdown of systems at low-load conditions. It can not only bring the advantage of better flexibility, but also utilize more energy and convert it into hydrogen. Finally, the hydrogen production under the proposed strategy is 43.84 kg higher than that of the traditional control strategy, as shown in TABLE II.

In conclusion, the proposed control strategy has the advantages of better flexibility, a wide load range, and higher PF compared to the traditional control strategy. And the production will not be a shortcoming under certain conditions, which include long-term low-load operating conditions.

### C. Cluster Characteristics of a Multi-stack P2H Load

In section IV.B, we have discussed the difference between the control strategy based on the proposed framework and the traditional control strategy based on a two-stack P2H load and reveal the tradeoff between production target and PF and security constraints. In this section, the proposed control strategy is applied to a multi-stack P2H load with ten stacks to explore the clustering characteristics.

We analyze operating characteristic to illustrate the typical different operating modes with different load levels. Here, five load levels of 10%, 20%, 30%, 50%, and 100% are discussed, the former three represent low load level scenarios, 50% represents a typical medium load level, and 100% is a high load level scenario. The average PF and daily operating modes are shown in TABLE III and Fig. 15.

TABLE III
THE AVERAGE PF UNDER DIFFERENT POWER INSTRUCTIONS

| Load level | 10% low | 20% low | 30% low | 50% medium | 100% high |
|---|---|---|---|---|---|
| PF | 0.9103 | 0.9116 | 0.9056 | 0.9058 | 0.9484 |

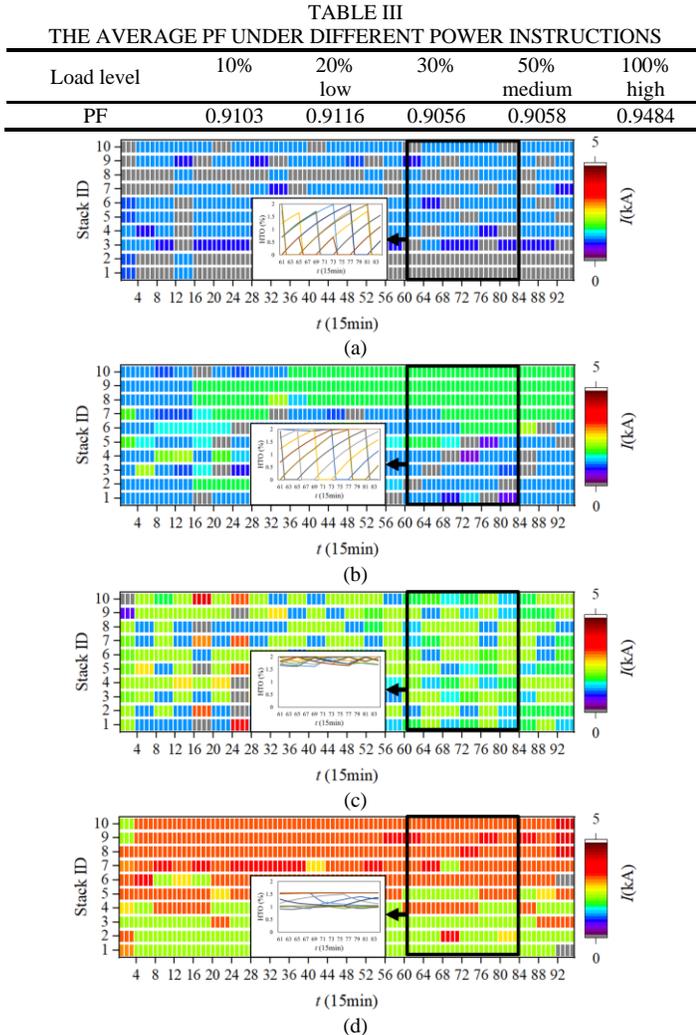

(a)

(b)

(c)

(d)

Fig. 15 The operating modes of a multi-stack P2H load under five typical constant power constructions: (a) 10%; (b) 20%; (c) 30%; (d) 50%

*1) Load level=10%.* In this scenario, a single stack's average load level can only attain 10%. Stacks cannot continue to operate at such a low load level due to the security constraint since HTO would attain its upper limit of 2% within around 3 hours. Therefore, Fig. 14(a) shows an average 4 hour-cycle of each stack to shut down to clear hydrogen impurity and restart. The cluster shows a shutdown-in-turn operating mode to ensure the security constraint. On this basis, stacks that are working obey the equal-split rule to pursue the maximal production, as shown in the box of t=60~84 min. In this scenario, the PF constraint is safe with an average value of 0.91.

*2) Load level=20%.* In this scenario, an operating mode that partial stacks take turns to shut down, and the others obey the equal-split rule can be seen at the cluster level, as shown in Fig. 14(b) and the box of t=60~84 min. Three stacks work on a constant load level of =1.5 kA with maximal production efficiency. Stacks 1~5 operate on a load level of <1 kA to meet the remaining power construction requirement and shut down every 4 hours to ensure security. The performance of Stacks 6 and 7 is between Stacks 1~5 and Stacks 8~10, and they are like a transitional phase. In this scenario, the PF constraint is safe with an average value of 0.91.

*3) Load level=30%.* In this scenario, Fig. 14(c) shows that based on the proposed control strategy, all stacks hold working status and adjust load power periodically in a cycle of around 4 hours with both PF and HTO constraints, as shown in the box of t=60~84 min. The average load level of a single stack is 30%. However, the equal-split control strategy cannot meet PF constraints, so an unequal current control strategy is adopted within [1,2] kA to clamp the average PF above 0.9. At the same time, the dynamic current adjustment is essential to limit the operating time of a stack at the low load level (1 kA) not exceeding 3 hour to ensure HTO constraint.

*4) Load level=50%.* In this scenario, Fig. 14(d) shows at each period, half of ten stacks work at around 3.5 kA, and other half works between [1.5,2] kA to clamp the average PF above 0.9. The current dispatch obeys the unequal-split rule. The model prediction control makes sure HTO intervals of the majority of separators are steady in [1,1.6]%. It represents a typical operation mode of a multi-stack P2H load at the medium load level.

*5) Load level=100%.* Finally, at the high load level, stacks obey same optimal control strategy that the equal-split rule as the traditional control strategy. Both PF and HTO constraints are slack, and all stacks constantly work without status and load power switch.

Finally, five corresponding operating modes can be concluded. Different load levels are accompanied by different control constraints, as shown in TABLE IV.

TABLE IV
THE TYPICAL OPERATING MODES UNDER DIFFERENT POWER INSTRUCTIONS

| Load level | HTO | PF | Modes description |
|---|---|---|---|
| 10% | √ | | status switch in-turn, current equal-split |
| 20% | √ | | partial status switch in-turn, current equal-split |
| 30% | √ | √ | no status switch, current adjust periodically |
| 50% | | √ | no status switch, current unequal-split |
| 100% | | | no status switch, current equal-split |

√This constraint has been triggered.

To verify the effectiveness of the proposed control framework, the time cost of the hour-ahead robust MPC and the real-time current correction are tested on a 10-stack P2H load.

TABLE V concludes the average time cost for determining a one-interval control strategy, i.e., 1-hour for the hour-ahead control and 15-minute for the real-time control. It shows that the proposed control framework can guarantee the control tractability with the premise of accuracy and robustness.

TABLE V
THE TIME COST OF THE PROPOSED CONTROL FRAMEWORK

|  | Hour-ahead robust MPC | Real-time current correction |
|---|---|---|
| Time cost (s) | 60.28 | 1.47 |

## Conclusion

This paper captures all-condition PF characteristics of multi-stack P2H loads and proposes a multi-timescale load control framework for the PF-constrained integration in volatile peak shaving conditions. For an industrial thyristor-based PCI, PF shows a nonmonotonic tendency that declines first and rises in the working interval of P2H, which leads to a PF-constrained infeasible region. The contrary monotonicity of PF compared with the efficiency results in the control tradeoff between the production target and the PF constraint. Case studies on a two-stack P2H load presents an unequal-split control strategy that is different from the traditional production-oriented optimal control strategy. The adoption of PF and security constraints in the control framework is verified to effectively improve the grid-side performance and the overall hydrogen production simultaneously when following the volatile power instructions. Furthermore, the operating characteristic at the cluster level is analyzed, and five operational modes are classified: status switch in-turn and current equal-split mode at low load level, current adjust periodically at medium load level, and current equal-split at high load level, which is referred for the optimal control in a realistic multi-stack P2H station.